\documentclass[prc,final]{revtex4}

\usepackage{graphicx}

\newcommand{\be}{\begin{equation}}
\newcommand{\ee}[1]{\label{#1} \end{equation}}
\newcommand{\ba}{\begin{eqnarray}}
\newcommand{\ea}[1]{\label{#1} \end{eqnarray}}
\newcommand{\nl}{\nonumber \\}

\newcommand{\exv}[1]{{\left\langle #1 \right\rangle}}

\begin{document}

\title{{\bf The hadronization line in stringy matter}}

\author{Tam\'as S. Bir\'o
}
\affiliation{
 KFKI Research Institute for Particle and Nuclear Physics Budapest
}
\author{Jean Cleymans
}
\affiliation{
Department of Physics,
 University of Cape Town, South Africa
}

\date{\today}

\begin{abstract}
The equation of state of the string model with linear strings 
comes close to describing the lattice QCD results and  allows
for the $E/N\approx 6T_0 = 1$ GeV relation 
found in phenomenological statistical model. 
The $E/N$ value is derived from the zero pressure condition in quark matter
and is a fairly general result.
The baryochemical potential dependence of the hadron gas  can be met if it
is re-interpreted in the framework of an additive quark model.
The conclusion  is reached that stringy models explain the $E/N=6T_0$
relation naturally and independently of the value of the string tension.
\end{abstract}

\maketitle

%
\vspace{7mm}
\section{Introduction}

The phenomenological success of the statistical hadronization model has been emphasized,
questioned and criticised repeatedly over the years.  
All experimental results  of heavy-ion collisions on particle yields at 
energies ranging from SIS to RHIC are consistent with results falling  in a narrow
stripe in the parameter space of temperature and baryochemical potential.
There are several descriptions of this so called chemical freeze-out curve,
for recent papers on this topic see \cite{STATMOD,FR1,FR2,FREEZE}.

In this paper we intend to explain the rather high value of the  
energy per particle,
$E/N \approx 1$ GeV for a system of  quarks and gluons with an
equation of state containing a term depending on the color density.
As a matter of fact the convex shape of this 
curve (cf. Fig.\ref{FIG_APPEND}) is in agreement with
several thermodynamic
approaches incorporating fermions and bosons as ideal gases.
The main puzzle is the quantitative value, since 1 GeV is six times higher than the
associated hadronization temperature, $T_0 \approx 167$ MeV. 
We aim to understand, how this value can be obtained
starting from a massless quark-gluon plasma by adding a color-density term.

On the hadronic side, considering massive matter, elementary 
non-relativistic thermodynamics leads to
$E/N=m+3T/2$. With the known values cited above this requires
an average mass in the range of the rho meson mass, $m  \approx 750$ MeV.
This value satisfies $m \gg T$, so the non-relativistic approximation
turns out to be acceptable.

However,  such a high mass on the quark matter side cannot be a constituent
mass, it can only be the result of strong interactions. 
A well known example for treating this interaction as a mean field,
the original MIT bag model, with 
$p=\kappa T^4-B$, $e=3\kappa T^4+B$ and $n\approx \kappa T^3$
pressure, energy density and particle density, respectively, would allow for
$E/N = e/n \le 4T$ only, upon $p \ge 0$.
Moreover the equation of state of the model quark matter
has to be in accordance to results obtained 
from lattice QCD, the only model independent non-perturbative
approach from field theory to the equation of state at present.
We present a simple thermodynamical model of massless quarks and gluons whose
interactions are
described by a free energy contribution motivated by strings.
This stringy matter will be studied at vanishing pressure for the hadronization curve
and at high temperature for the lattice equation of state.
It is an interesting conclusion that such stringy models do explain the $E/N=6T_0$
relation easily and, remarkably,  independent of the value of the string tension.

\section{Stringy thermodynamics}

There are several ways to treat corrections to an ideal gas equation of state (eos).
Each model identifies a physical picture in which the leading order interactions
are calculated. The lattice QCD and pure Yang Mills eos approach an effective
massless ideal gas limit at high temperatures, but it deviates vastly from it around
the color deconfinement temperature. In this paper we test the string model picture
for an interacting gluonic plasma, introduced in \cite{BiroShanenko}.

We assume that a however decreasing proportion of color charges are still connected
by strings at high temperature, above the crossover value to deconfinement $T > T_c$.
This assumption is supported by some simple and general properties of the thermal
distribution of relative momentum squared between pairs of massless particles,
\be
 P(Q^2) = \exv{\delta\left(Q^2-2E_1E_2(1-\cos\vartheta)\right)} =
 \frac{\int_0^\infty\limits\int_{Q^2/4E_1}^{\infty}\limits E_1E_2 f(E_1) f(E_2)}{4\int_0^\infty\limits\int_{0}^{\infty}\limits E_1^2E_2^2 f(E_1) f(E_2)}.
\ee{PQ2}
This distribution has a non-perturbative contribution to the equation of state
stemming from relative momenta below a fixed value, 
($Q^2 < \Lambda^2$ with say $\Lambda\approx 1$ GeV):
\be
 F = \int_0^{\Lambda^2}\limits P(Q^2)  \, dQ^2. 
\ee{IPQ2}
This quantity, not having another energy scale than the temperature in case of massless
particles is given by a scaled integral
\be
 F = \int_0^{\Lambda^2/T^2}\limits f(x) \, dx .
\ee{SCALED}
For high temperatures $T \gg \Lambda$ this proportion is approximately given by
$F \approx f(0) \: \Lambda^2/T^2$, for low temperature, $T \ll \Lambda$
by $F \approx 1$ due to normalization. Since 
\be
f(0) =  \exv{\frac{T}{2E}}^2
\ee{NONZERO} 
is non-zero (cf. eq. \ref{PQ2}), 
there is always a
non-perturbative contribution to the equation of state at any temperature, it just
becomes subleading order in the pressure. Nevertheless in the interaction measure,
$(e-3p)$, such an ${\cal O}(T^2\Lambda^2)$ term mixes to the leading order.


\subsection{Equation of state with strings}

A general class of eos is obtained by taking into account a term
in the free energy proportional to a fractional power of 
the density.
Here we consider a ''color density'', a weighted sum of the number of particles:
\be
 c = \sum n_i c_i.
\ee{COLOR}
Colorless objects do not pull strings, 
nor do they take part in screening (ending) them.
Furthermore different color charges like quarks or gluons 
may have different effective string constants \cite{COLORROPE};
this effect is taken into account in 
the factors $c_i$.

The contribution to strings of an average length of
\be
 \langle \ell \rangle  \propto c^{-\gamma}
\ee{AVLENGTH}
shall be multiplied by the density,
with $\gamma$ being a fractional power between zero and one. 
For straight strings in  three dimension $\gamma=1/3$.

Since we consider density dependent modifications of the eos at a given temperature,
we shall use the free energy density, 
denoted $f$, 
as the fundamental potential. We consider
\be
 f = \tilde{f}(n_i,T) + \frac{A}{1-\gamma} \, c^{\: 1-\gamma}.
\ee{GENFREE}
Here the coefficient $A$ comprises eventual average geometrical shape factors besides
the string tension, so it cannot be taken from the Regge slope of meson resonances
directly. We shall rather fit it to 
the
lattice eos later.
The  $\tilde{f}$ is a general free energy density, for describing a QGP to be specified later.

The chemical potentials associated to the component $i$ are given by the derivatives
with respect to $n_i$:
\be
 \mu_i = \tilde{\mu}_i +A c^{-\gamma} \, c_i
\ee{CHEMi}
where $\tilde{\mu}_i=\partial\tilde{f}/\partial n_i$ follows from $\tilde{f}$. 
We use so far the additive form of eq.(\ref{GENFREE}).
The chemical equilibrium establishes, if possible, at constant values of this
chemical potentials. These values are determined by the conserved charges of the
components. For the sake of simplicity we consider here the baryon charge only
(zero for gluons, $1/3$ and $-1/3$ for quarks and antiquarks, respectively), but
further quantities may be introduced into this scheme easily. From the equations
\be
 \mu_i = q_i \mu_B
\ee{EQchemi}
the corresponding number densities can be expressed
\be
 n_i^{{\rm eq}} = \nu_i(T,\mu_B;q_i,c_i).
\ee{NUMdensi}
In this chemical equilibrium situation the color density defined in eq.(\ref{COLOR})
takes its equilibrium value expressed by the $\nu_i$-s.

While the real solution of this system of equations can be involved in 
the 
general case,
in some particular systems, e.g. for a massless ideal gas with strings in the
Boltzmann approximation, it can be given in analytic
form (see later). A remarkable general property, however, can be derived without
the explicit form of this solution. This property related to the hadronization of
the QGP is the energy per particle at the edge of the stability: when the pressure
vanishes. For this purpose we obtain the entropy density,
\be
s = -\frac{\partial f}{\partial T} = \tilde{s},
\ee{GENENT}
the pressure
\be
p = \sum \mu_i n_i -f = \tilde{p}-\frac{\gamma A}{1-\gamma}\, c^{\: 1-\gamma}
\ee{GENPRESS}
and the energy density
\be
e = f+Ts = \tilde{e} + \frac{A}{1-\gamma}\, c^{\: 1-\gamma}.
\ee{GENERG}
At the endpoint of the mechanical stability $p=p_0=0$ and therefore
\be
 \tilde{p}_0 = \frac{\gamma A}{1-\gamma}\, c^{\: 1-\gamma}.
\ee{MECHEND}
From this the color weighted density can be obtained at this point,
\be
 c_0 = \left(\frac{1-\gamma}{ A\gamma} \: \tilde{p}_0 \right)^{\frac{1}{1-\gamma}},
\ee{ENDCOLOR}
and the energy density is expressed as
\be
 e_0 = \tilde{e}_0 + \tilde{p}_0 / \gamma.
\ee{ENDERG}
This is a remarkable result. For  massless constituents only 
$\tilde{p}=T^4\phi(\mu_B/T)$ and therefore $\tilde{e}=3\tilde{p}$.
For straight strings $\gamma=1/3$, so we arrive at $e_0=6\tilde{p}_0$.
As long as the Boltzmann approximation is applicable, for an ideal gas mixture
$\tilde{p}=nT$ with $n=\sum n_i$ and one concludes that
\be
 \frac{E}{N} = e_0 / n = 6 T_0.
\ee{GENEN}
With $T_0=167$ MeV fitted to hadronization data (and predicted by lattice QCD as
the crossover temperature) one obtains $E/N=1$ GeV at this point. 
This %
derivation applies to the 
quark-gluon  side of the hadronization curve.


\subsection{Chemical equilibrium with strings}

From eqs.(\ref{CHEMi}) and (\ref{EQchemi}) follows that the equilibrium number densities
in general satisfy
\be
 n_i^{{\rm eq}} = n_{i,{\rm A=0}}^{{\rm eq}}(T,q_i\mu_B-Ac^{-\gamma}c_i).
\ee{NUMi}
Summing 
with the color weight factors, $c_i$ we arrive at an implicit equation
for the effective color (string source) density:
\be
 c = \sum c_i \: n_{i,{\rm A=0}}^{{\rm eq}}(T,q_i\mu_B-Ac^{-\gamma}c_i).
\ee{COLOReff}
Its solution in general depends on the temperature $T$, baryochemical potential $\mu_B$
and the specific color and baryon charges $c_i$ and $q_i$.

Since this implicit equation contains the unknown, $c$, on its right hand side as
a given correction to the chemical potential only, we are able to deduce that there
is no solution beyond a critical point. This point is characterized by the fact that
the derivative of eq.(\ref{COLOReff}) with respect to $c$ is also satisfied,
\be
 1 = \gamma A c^{-\gamma-1} \sum c_i^2 \: \frac{\partial n^{{\rm eq}}}{\partial \mu}
   (T,q_i\mu_B-Ac^{-\gamma}c_i).
\ee{DERIV}
We denote the color density in this endpoint where the chemical equilibrium
ceases to be possible by $c_E$.
A further question is how this endpoint of possible chemical equilibrium solutions
is related to the zero pressure boundary.

In the particular case of Boltzmann-like dependence on the chemical potential
$\partial n_i/\partial\mu = n_i/T$ (i.e. the exponential of $\mu/T$ is a factor
in the number density).
We note that for a Boltzmann system with all $c_i$'s equal to either one or zero 
$c_i^2=c_i$ and therefore the ratio of eqs.(\ref{DERIV}) and (\ref{COLOReff}) gives
a simple condition for the critical color density $c=c_E$:
\be
 \gamma \: \frac{A}{T} \: c^{-\gamma}_{E}  \: = \: 1.
\ee{SIMPLEcrit}
In the same approximation one can obtain an analytic
solution to the chemical equilibrium problem. We get
\be
 c = \left(\sum_{c_i=1}\limits \chi_i(T)e^{q_i\mu_B/T} \right) \, e^{-Ac^{-\gamma}/T}
\ee{BOcolor}
with
\be
 \chi_i(T) = n_{i, {\rm A=0}}^{{\rm eq}}(T,0).
\ee{ChiDEF}
The sum in the brackets is a $c$-independent factor, we denote it by 
$\varphi(T,\mu_B)=T^3a^3(\mu_B,T)$.
The solution of eq.(\ref{BOcolor}) is given by 
\be
 z = \frac{\gamma Ac^{-\gamma}}{T} = -W(-\frac{\gamma A}{T} \varphi^{-\gamma})
\ee{SOLUT}
with $W(x)$ being Lambert's W-function defining the solution of the transcendental equation
$We^W=x$. The form occurring in eq.(\ref{SOLUT}), $-W(-x)$ has a real solution for positive $x$
values only if $x>1/e$. The argument $x=e^{-1}$ is the endpoint of stringy
thermodynamics, the endpoint temperature is connected to the effective string tension
$A$ as 
\be
 \gamma \frac{A}{T_E} \varphi(T_E,\mu_B)^{-\gamma} = e^{-1}.
\ee{ENDPOINTA}

For considering massless particles as sources and endings of strings the factor
$\varphi=a^3T^3$ has a special form: $a(\mu_B,T)$ can be a function of the ratio $\mu_B/T$ only.
At vanishing baryochemical potential, $\mu_B=0$, $a$ is a constant. In this case
eq.(\ref{ENDPOINTA}) can be resolved for the endpoint temperature:
\be
  T_E = \left(\gamma A e a^{-3\gamma} \right)^{\frac{1}{3\gamma + 1}}.
\ee{TEEND}
In particular for $\gamma=1/3$ it is $T_E=\sqrt{Ae/3a}$ and $a=a(\mu_B/T_E)$.
This implicitely determines the endline $\mu_B(T_E)$.

\section{Massless Boltzmann gas with strings at $\mu_B=0$}

In the followings we investigate the one component massless Boltzmann gas at vanishing net baryon
density. In this case $a$ is constant and the pressure and energy density are
expressed easily with help of explicit chemical equilibrium solution  
\be
 c = a^3T^3 \left(\frac{-W(-x)}{x} \right)^{-1/\gamma}
\ee{cCHEMEQ}
with
\be
 x = e^{-1} \left(\frac{T_E}{T}\right)^{3\gamma+1}.
\ee{XARG}
Since for ideal Boltzmann gases the Boyle-Mariotte law holds, $\tilde{p}=cT$,
and for vanishing mass $\tilde{e}=3\tilde{p}$ the total pressure is reconstructed
as (cf. eq.\ref{GENPRESS}):
\be
 p = cT \left( 1 - \frac{z}{1-\gamma}\right)
\ee{BOLPRESS}
where $z$ is taken from the solution (\ref{SOLUT}).
The energy density is given by
\be
 e = cT \left( 3 + \frac{z}{\gamma(1-\gamma)}\right)
\ee{BOLENERG}

\subsection{High temperature asymptotics}

When comparing with lattice results the high temperature asymptotics of the stringy
massless Boltzmann eos described by the eqs.(\ref{SOLUT},\ref{BOLPRESS},\ref{BOLENERG})
implicitly is particularly important. In our model $T \rightarrow \infty$
is realized by the $x \rightarrow 0$ limit (we consider $\gamma \in (0,1)$ only).
The Lambert W-function has the series expansion $z=-W(-x)=x+x^2+\ldots$ and hence
the high temperature (low-x) color density behaves like
\be
 c = a^3T^3 \left( 1 -\frac{1}{\gamma} x + \ldots \right),
\ee{LOWxC}
the pressure like
\be
 p = a^3T^4 \left( 1 -\frac{1}{\gamma(1-\gamma)} x + \ldots \right)
\ee{LOWxPRESS}
and the energy density like
\be
 e = a^3T^4 \left( 3 +\frac{3\gamma-2}{\gamma(1-\gamma)} x + \ldots \right).
\ee{LOWxENERG}
The interaction measure is given by
\be
 e - 3p = a^3T^4 \left(\frac{3\gamma+1}{\gamma(1-\gamma)} x + \ldots \right).
\ee{LOWxIMES}
This analysis reveals that the high temperature asymptotics of the pressure is
$T^4$ as it should be; this fixes the value of $a^3$ by comparing to lattice eos.
(5.21 for 2+1 flavor QCD, 1.7 for pure SU(3)).

The power $\gamma$ can be fixed from the subleading terms (and leading term in the
interaction measure), $xT^4$. Using eq.(\ref{XARG}) such terms scale like $T^{3-3\gamma}$.
In order to agree with lattice eos results, i.e. $xT^4\sim T^2$ the only possible
choice we are left with is $\gamma=1/3$. 
This agrees with our statistical argument for leading non-perturbative effects
on the eos at $T > T_c$, and in particular favors the straight string
geometry in quark matter.

\subsection{The zero pressure point as reference}

Since on the lattice only positive pressure states can be achieved by Monte Carlo
techniques, the zero pressure point, $T_0$, is a more physical reference than the
endpoint of chemical equilibrium. This occurs at $z_0=1-\gamma=2/3$ and due to
$z=xe^z$ at $x_0=\frac{2}{3}e^{-2/3}\approx 0.34$. On the other hand
$x_0=e^{-1}(T_E/T_0)^2$ due to eq.(\ref{XARG}), so we arrive at
$T_0/T_E=(x_0e)^{-1/2}\approx 1.04$. In the stringy massless Boltzmann quark
matter approximation these two temperatures are quite close to each other.
Reference to the zero pressure point can be taken by substituting 
$x=x_0(T_0/T)^2 \approx 0.34 (T_0/T)^2$ in general.

\begin{figure}
\begin{center}
\includegraphics[width=0.35\textwidth,angle=-90]{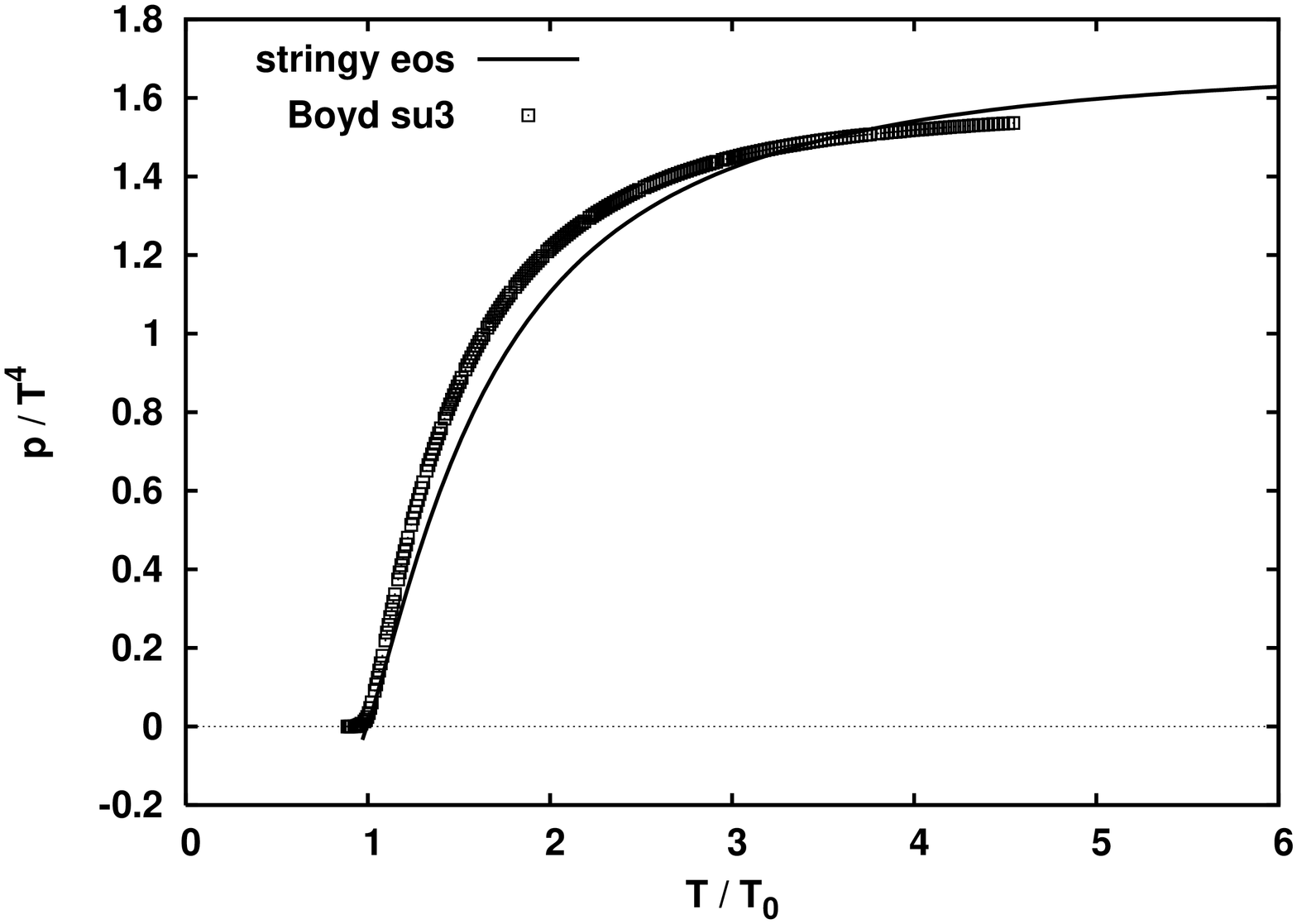}\includegraphics[width=0.35\textwidth,angle=-90]{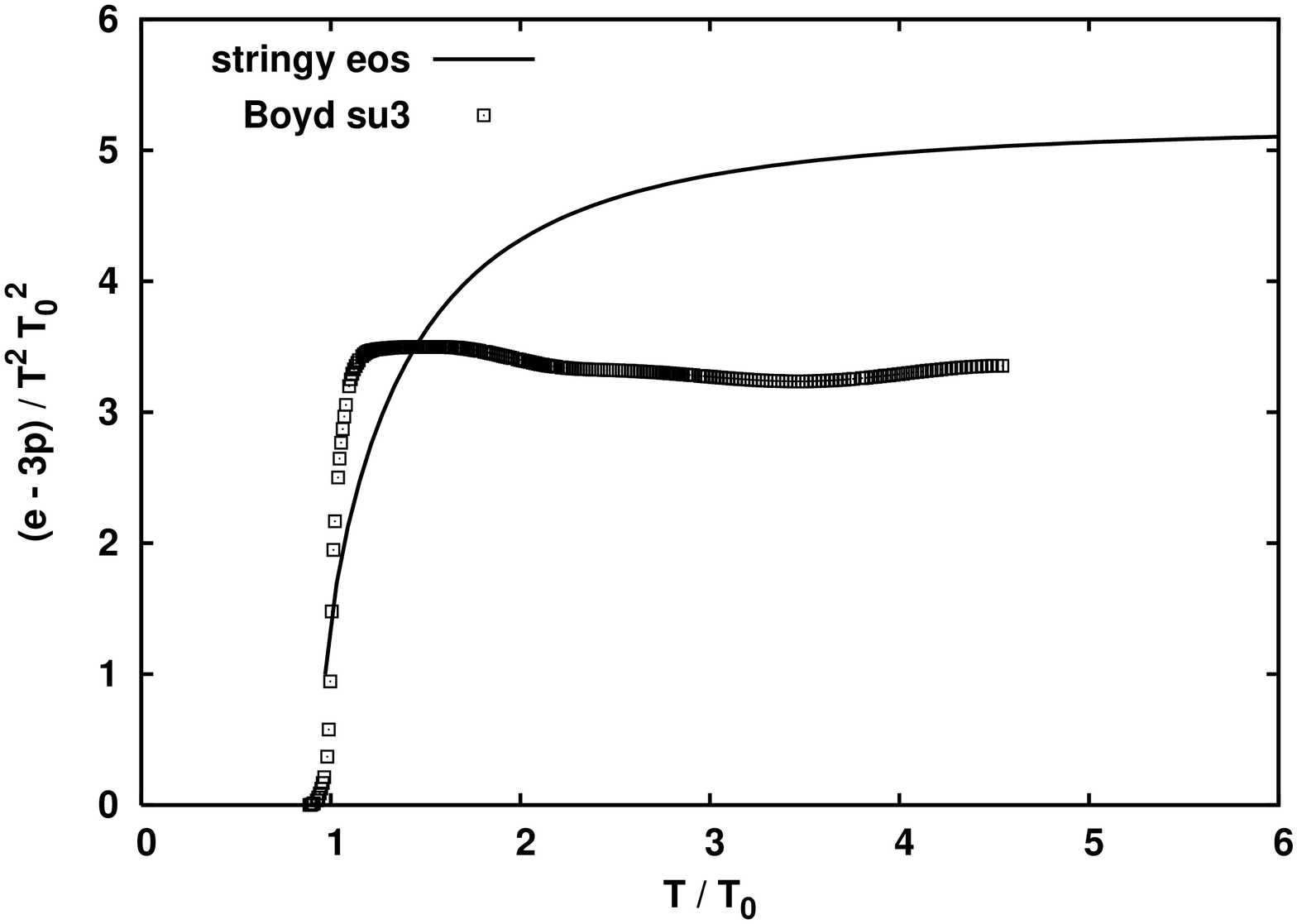}
\includegraphics[width=0.35\textwidth,angle=-90]{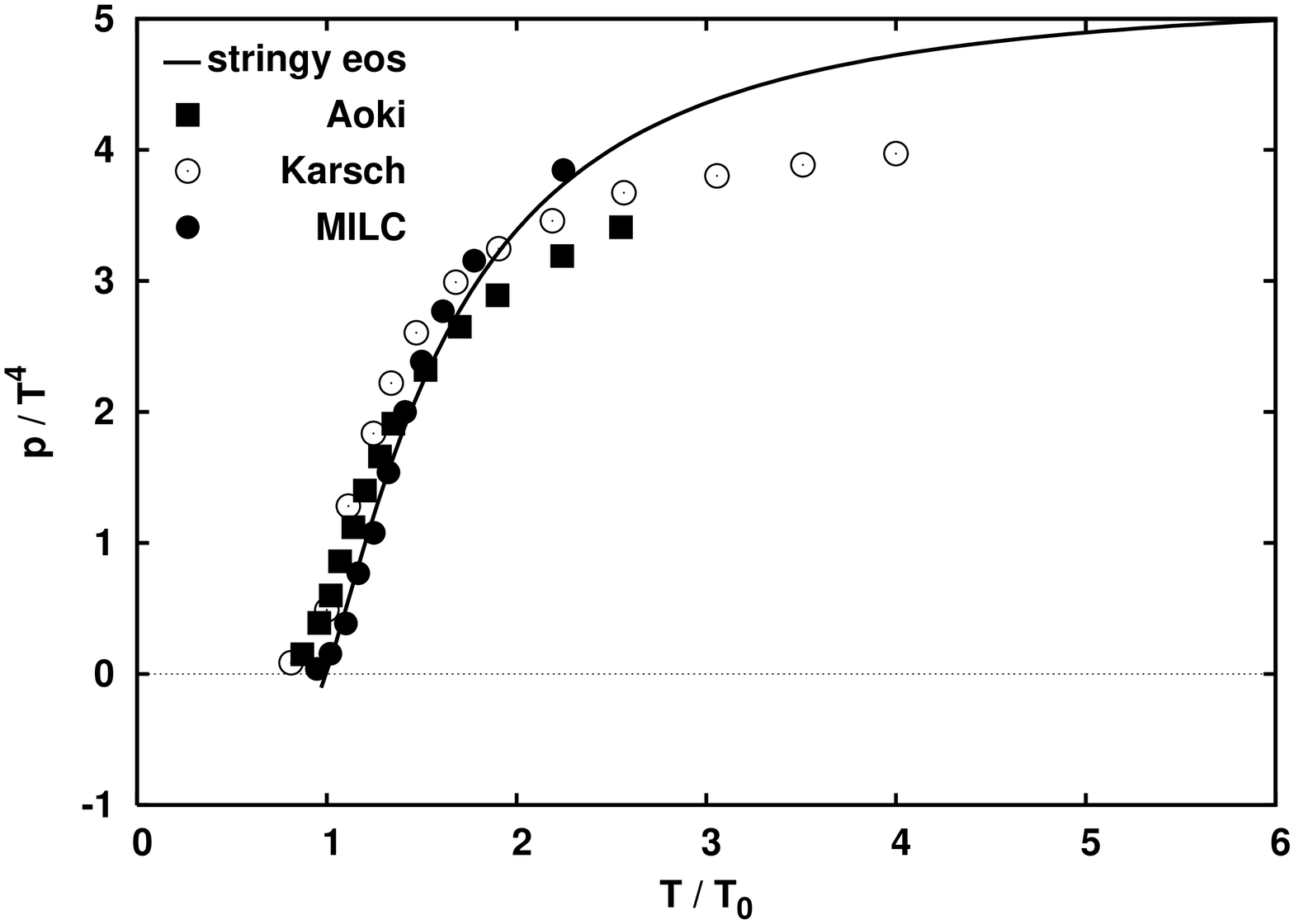}\includegraphics[width=0.35\textwidth,angle=-90]{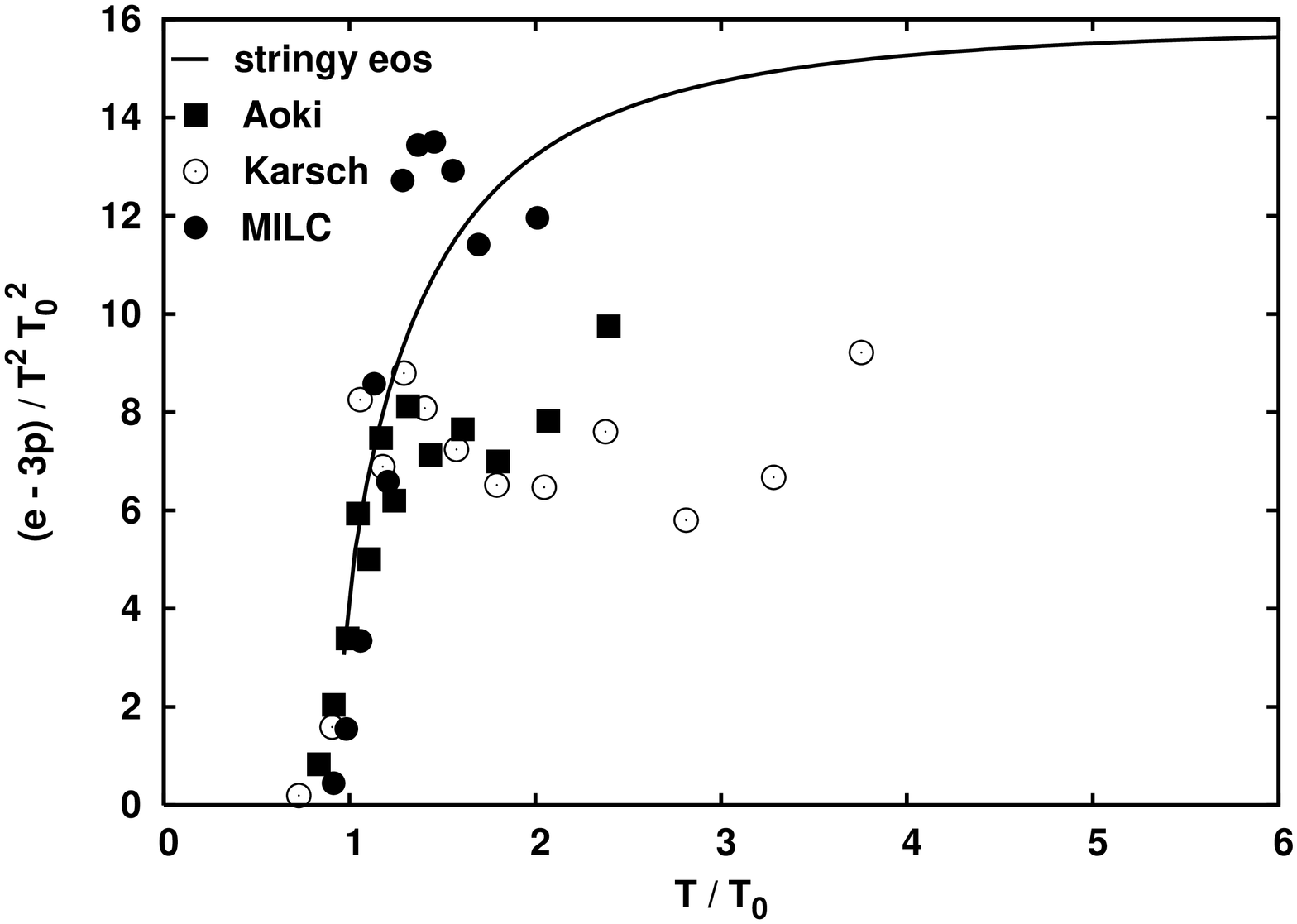}
\end{center}
\caption{\label{EOS} Equation of state from lattice and 
from the massless Boltzmann stringy model. Pure SU(3) gauge theory results
\cite{Karsch} to the top,
2+1 flavor QCD results \cite{BIELEFELD,FODOR,MILC} 
to the bottom. The scaled pressure (left sides) 
and interaction measure (right sides) curves are compared.
}
\end{figure}

With respect to $T_0$ the scaled pressure and interaction measure are given as
\ba
 \frac{p}{T^4} &=& \frac{x^3a^3}{z^3} \left(1-\frac{3z}{2} \right) \nl
 \frac{e-3p}{T_0^2T^2} &=& \frac{9a^3x^2}{z^2} x_0. 
\ea{ZEROEOS}
This reveals a factor of $9x_0 \approx 3.04$ between the high-temperature asymptotic
values of the scaled interaction measure $(e-3p)/T^2T_0^2$ and scaled pressure $p/T^4$.
Although this gives a qualitatively correct picture of the quark matter eos,
unfortunately for a quantitative comparison Monte Carlo data are still
spread over a remarkable range (cf. Fig.\ref{EOS}).


The whole temperature dependence of the stringy eos in Boltzmann approximation works
surprisingly well for the pressure, but
is
 only qualitatively correct for the interaction
measure. This we plot in Fig.\ref{EOS}. The lattice data show the most sudden jump
in the $T^2$-scaled interaction measure to a nearly constant value. For comparison
the stringy model has a somewhat milder rise. 
Quantitatively the constant value of the scaled interaction measure differs
from the lattice result in the pure SU(3) case \cite{Karsch} 
as well as probably also in the full QCD case \cite{BIELEFELD,FODOR,MILC}. 
The massless Boltzmann approximation is clearly an oversimplification.

\vspace{10mm}
\subsection{Degenerate Fermi gas at T=0}

Another analytically solvable case is represented by the degenerate Fermi gas
at zero temperature with stringy contributions to thermodynamics.
In this case, considering only one type of fermions with color charge $c_i=1$
the chemical potential fixes the Fermi energy. Let us consider the case of
massless fermions, than the Fermi momentum is also $\mu$ and the chemical
equilibrium relates the color density to this potential via
\be
 c = \frac{d}{6\pi^2} \left( \mu - Ac^{-1/3} \right)^3,
\ee{FERMIc}
with degeneracy factor $d=(2S+1)N_cN_f=12$ for light quark matter.

This equation is analytically solvable. Introducing the parameter
$B=(6\pi^2/d)^{1/3}$ we have a second order algebraic equation for the variable $c^{1/3}$:
\be
 B c^{1/3} + A c^{-1/3} = \mu. 
\ee{FERcQUAD}
The stable solution belongs to the higher value of $c$
(because if the lhs of eq.(\ref{FERcQUAD}) exceeds $\mu$ then chemical processes tend
to decrease $c$):
\be
 c = \left(\frac{\mu+\sqrt{\mu^2-4AB}}{2B} \right)^3.
\ee{FERcSOL}
It is easy to see that the critical value of the chemical potential is given by
\be
 \mu_E = 2\sqrt{AB},
\ee{FEREND}
for $\mu < \mu_E$ there is no equilibrium, the strings pull the matter infinitely.

For obtaining the pressure and energy density we use the corresponding expressions for
a degenerate Fermi gas:
\ba
 \tilde{p} &=& \frac{d}{24\pi^2} \left(\mu - Ac^{-1/3} \right)^4  \nl
 \tilde{e} &=& 3 \tilde{p}
\ea{FERDEG}
Upon using eq.(\ref{FERMIc}) this leads to the total pressure
\be
 p = \frac{1}{4} \mu c - \frac{3}{4} A c^{2/3},
\ee{FERPRES}
and to the energy density
\be
 e = \frac{3}{4} \mu c + \frac{3}{4} A c^{2/3}.
\ee{FERENER}
At zero pressure, $p=0$, the energy per particle becomes $e/n=\mu$ and the value of
the chemical potential at this point can be obtained from eq.(\ref{FERPRES}) as satisfying
$\mu_0c_0=3Ac_0^{2/3}$. Together with eq.(\ref{FERMIc}) this leads to
\be
 \mu_0 = 3\sqrt{AB/2} = \frac{3}{2\sqrt{2}} \mu_E \approx 1.06 \mu_E
\ee{FERMuNULL}
at the color density $c_0=(2A/B)^{3/2}$. If this meets the $E/N=1$ GeV line, then
at $T=0$ $\mu_0=1$ GeV.

\begin{figure}
\begin{center}
\includegraphics[width=0.4\textwidth,angle=-90]{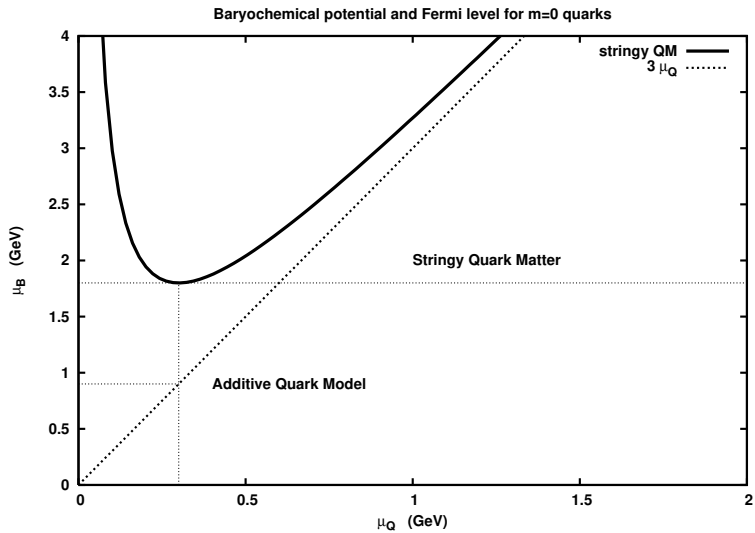}
\end{center}
\caption{ \label{FIG_MUBAR}
  The baryochemical potential as a function of the Fermi momentum of quarks at $T=0$
  in the stringy QGP. The minimum point represents the endpoint of the stringy model,
  the corresponding additive quark model value is the half of it.
}
\end{figure}

Since the endpoint and the zero pressure points are close to each other both at
zero baryochemical potential and at zero temperature, we conjecture, that this is the
case all over the hadronization line. The baryochemical potential at the endpoint
of the stringy QGP is given by the minimum value of the curve
\be
 \mu_B = \frac{3AB}{k_F} + 3k_F
\ee{BARCURVE}
with $k_F$ being the Fermi momentum (for massless fermions it is also the Fermi energy,
i.e. the chemical potential of the quarks). Its minimum is located at $k_0=\sqrt{AB}$,
the value being $\mu_{B}^E=6\sqrt{AB}$. However, regarding the hadrons fitted in the
Statistical Model as objects made in the constituent (additive) quark model, one has to
consider $\mu_B^A=3k_0=3\sqrt{AB}$. This value is exactly the half of the previous one.
Fig.\ref{FIG_MUBAR} plots this relation.

\begin{figure}
\begin{center}
\includegraphics[width=0.4\textwidth,angle=-90]{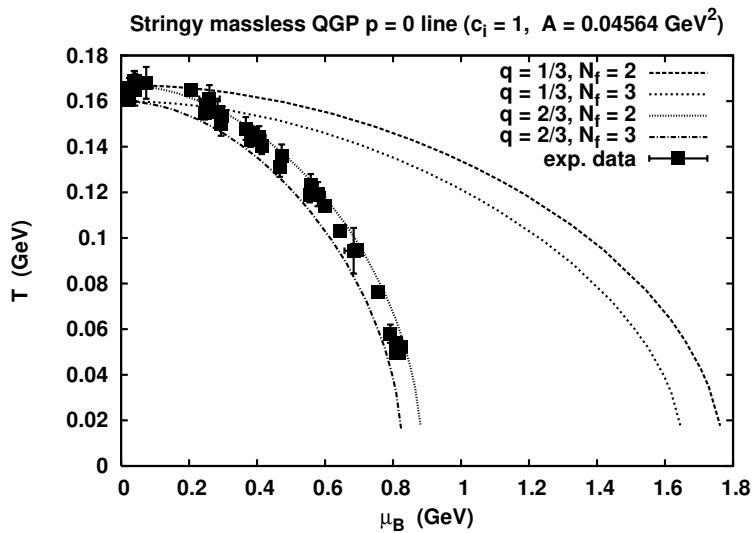}
\end{center}
\caption{ \label{FIG_APPEND}
  The zero pressure line of the stringy, massless QGP (labeled by $q=1/3$) 
  with 2 and 3 quark flavors. 
  For comparison the Statistical Model results (boxes), fit by $T=0.167-0.139\mu_B^2-0.053\mu_B^4$, 
  Ref.\cite{FREEZE} are indicated. We also plotted the stringy QGP hadronization
  lines with halved baryochemical potentials roughly corresponding to an additive quark
  model interpretation of hadrons (labeled by $q=2/3$).
}
\end{figure}

\begin{figure}
\begin{center}
\includegraphics[width=0.4\textwidth,angle=-90]{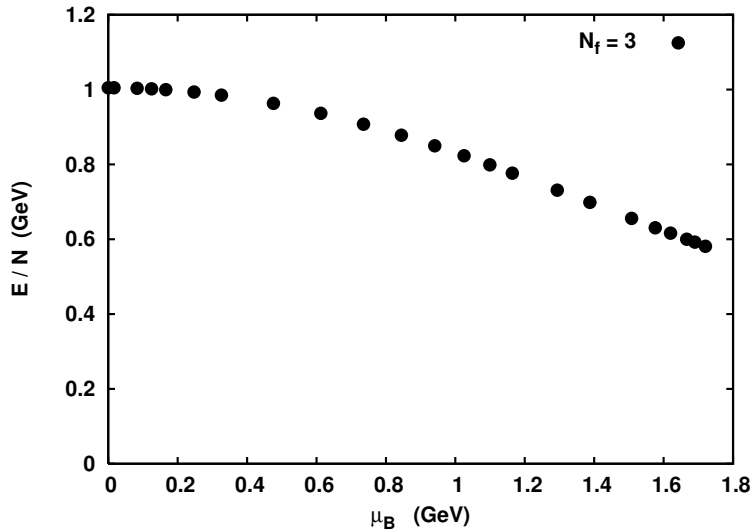}
\end{center}
\caption{ \label{FIG_EPERN}
  The energy per particle for a stringy, massless QGP with 3 light flavors.
  Until the value $\mu_B  \approx 0.8$ GeV the $E/N$
  ratio does not change more than 10 per cent.
}
\end{figure}

\vspace{3mm}
The corresponding hadronization curves are plotted in Fig.\ref{FIG_APPEND}.
The $p=0$ line (which is close to the endline of the stringy thermodynamics within
a few per cent) is indicated for a quark gluon plasma with massless gluons and
$N_f=2$ and $N_f=3$ massless quark and antiquark flavors, respectively. For comparison
the Statistical Model results are indicated by filled boxes. Following our previous
comment about the additive quark model interpretation of the stringy QGP results,
the same hadronization lines are also plotted as functions of three times the quark
Fermi level instead of the baryochemical potential. In this second case the
coverage with the hadronic fit results is intriguing.
In Fig.\ref{FIG_EPERN} we plot the energy per particle along the hadronization line;
it is roughly constant for moderately low values of the baryochemical potential, but
then decreases somewhat.

\vspace{5mm}

Summarizing the stringy eos with $\gamma=1/3$ comes close to describing the
$T>T_0$ lattice QCD results and at the same time offers an explanation
for the $E/N\approx 6T_0 = 1$ GeV hadronization condition found in phenomenological
statistical model. While the massless Boltzmann approximation should be
dropped in the view of lattice eos data, the $E/N$ value at zero pressure
is a much more general result; it hopefully survives as a possible physical
picture for the quark matter side at hadronization in relativistic heavy ion
collisions. 
Admittedly there is no reason for the $E/N$ ratios being equal in the hadronic and
quark matter, but it is not easy to imagine, how
this value could be doubled from $3T_0$ to $6T_0$ at the hadronization.
The baryochemical potential dependence of the hadronic side fit
by the Statistical Model can be met if the chemical potential is re-interpreted
in the framework of an additive quark model (parameterized by the half of the
stringy QGP value).

\vspace{7mm}
{\bf Acknowledgment}

We acknowledge enlightening discussions with Peter Levai. This work has been
supported by the Hungarian Scientific Research Fund (OTKA K49466), by
a common project of OTKA and the National Office for Research and Technology 
(OTKA-NKTH 68108) and by the Hungarian-South-African 
Intergovernmental S\&T Cooperation Programme.
This work has been completed during T.S.Biro's participation in the workshop
"New Frontiers in QCD 2008" held at the Yukawa Institute, Kyoto, Japan.



\begin{thebibliography}{xxxxx}



\bibitem{STATMOD} 
J.~Cleymans, H.~Satz, Z. Phys. {\bf C 57}, 135, 1993;
P.~Braun-Munzinger, J.~Stachel, J.~P.~Wessels, N.~Xu, 
Phys.~Lett.~{\bf B 344}, 43, 1995; {\bf B 365}, 1, 1996;
P.~Braun-Munzinger, I.~Heppe, J.~Stachel, 
Phys.~Lett.~{\bf B 465}, 15, 1999;
P.~Braun-Munzinger, D.~Magestro, K.~Redlich, J.~Stachel, 
Phys.~Lett.~{\bf B 518}, 41, 2001;
J.~Cleymans, H.~Oeschler, K.~Redlich, S.~Wheaton, 
J.~Phys.~{\bf G 32}, S165, 2006.


\bibitem{FR1} 
  J.~Cleymans, K.~Redlich,
  Phys. Rev. Lett. {\bf 81}, 5284, 1998.

\bibitem{FR2} 
  J.~Cleymans, K.~Redlich,
  Phys. Rev. C {\bf 60}, 054908, 1999.

\bibitem{FREEZE} 
  J.~Cleymans, H.~Oeschler, K.~Redlich, S.~Wheaton,
  Phys. Rev. C {\bf 73}, 034905, 2006.


\bibitem{BiroShanenko} 
 T.~S.~Biro, A.~A.~Shanenko, V.~D.~Toneev,
 {\em Toward Thermodynamic Consistency of Quasiparticle Picture},
 Phys. Atomic Nuclei, {\bf 66}, 982, 2003.


\bibitem{COLORROPE} 
T.~S.~Biro, H.~B.~Nielsen, J.~Knoll,
Nucl.~Phys.~{\bf B 245}, 449, 1984.


\bibitem{Karsch} 
G.~Boyd, J.~Engels, F.~Karsch, E.~Laermann, C.~Legeland, M.~Lutgemeier, B.~Petersson,
{ Nucl.~Phys.}~{\bf B 469}, 419, 1996.

\bibitem{BIELEFELD} 
F.~Karsch, E.~Laermann, A.~Peikert, 
{ Phys.~Lett.}~{\bf B 478} 447, 2000.


\bibitem{FODOR} 
Y.~Aoki, Z.~Fodor, S.~D.~Katz, K.~K.~Szabo,
JHEP {\bf 0601}, 089, 2006; { Phys.~Lett.}~{\bf B643}, 46-54, 2006.

\bibitem{MILC} 
C.~Bernard, T.~Burch, C.~DeTar, S.~Gottlieb, L.~Levkova, U.~M.~Heller, J.~E.~Hetrick, R.~Sugar,
D.~Toussaint,
Phys.~Rev.~{\bf D 75}, 094505, 2007.

\end{thebibliography}
\end{document}